# Automatic Breast Lesion Detection in Ultrafast DCE-MRI Using Deep Learning

**Fazael Ayatollahi,[a] Shahriar B. Shokouhi,[b] Ritse M. Mann,[c] Jonas Teuwen,[d]**

[a] fazael.ayatollahi@radboudumc.nl,
 Electrical Engineering Department, Iran University of Science and Technology (IUST), Tehran, Iran
 Department of Radiology and Nuclear Medicine, Radboud University Medical Center, Nijmegen, the Netherlands
[b] bshokouhi@iust.ac.ir,
 Electrical Engineering Department, Iran University of Science and Technology (IUST), Tehran, Iran
[c] Ritse.Mann@radboudumc.nl,
 Department of Radiology and Nuclear Medicine, Radboud University Medical Center, Nijmegen, the Netherlands
[d] Jonas.Teuwen@radboudumc.nl,
 Department of Radiology and Nuclear Medicine, Radboud University Medical Center, Nijmegen, the Netherlands
 Department of Radiation Oncology, Netherlands Cancer Institute, Amsterdam, the Netherlands

**Corresponding Author:**
Fazael Ayatollahi
E-mail: fazael.ayatollahi@radboudumc.nl

**Abstract**

**Purpose:** We propose a deep learning-based computer-aided detection (CADe) method to detect breast lesions in ultrafast DCE-MRI sequences. This method uses both the three-dimensional spatial information and temporal information obtained from the early-phase of the dynamic acquisition.

**Methods:** The proposed CADe method, based on a modified 3D RetinaNet model, operates on ultrafast T1 weighted sequences, which are preprocessed for motion compensation, temporal normalization, and are cropped before passing into the model. The model is optimized to enable the detection of relatively small breast lesions in a screening setting, focusing on detection of lesions that are harder to differentiate from confounding structures inside the breast.

**Results:** The method was developed based on a dataset consisting of 489 ultrafast MRI studies obtained from 462 patients containing a total of 572 lesions (365 malignant, 207 benign) and achieved a detection rate, sensitivity, and detection rate of benign lesions of 0.90 (0.876-0.934), 0.95 (0.934-0.980), and 0.81 (0.751-0.871) at 4 false positives per normal breast with 10-fold cross-testing, respectively.

**Conclusions:** The deep learning architecture used for the proposed CADe application can efficiently detect benign and malignant lesions on ultrafast DCE-MRI. Furthermore, utilizing the less visible hard-to detect-lesions in training improves the learning process and, subsequently, detection of malignant breast lesions.

# 1 Introduction

Breast cancer is the most common cancer among women and an important cause of cancer-related death worldwide [1,2]. Mammography has played an important role in reducing breast cancer mortality during the past decades. Many countries have implemented population-based screening programs using mammography due to its low cost and the short acquisition time [3]. However, dynamic contrast-enhanced magnetic resonance imaging (DCE-MRI) is a 4D imaging modality that provides higher sensitivity than mammography. In women at increased risk, mammography alone proved to be insufficient, and in these women breast MRI has been added to most screening programs. In practice, by far, most cancers in these women are currently detected by screening breast MRI. Screening for breast cancer follows an intensified scheme in women at higher than average risk for the development of breast cancer. According to the American cancer society guidelines [4] women with a lifetime risk of 20% or more should be screened with annual breast MRI, which is supplemented by mammography from the age of 30. In high-risk screening, the sensitivity of MRI is about double that of mammography [5]. In fact, the added detection of cancers detected by mammography alone accounts to less than 10% of all cases and is largely limited to DCIS and low-grade lesions [6]. Mammography is also proven to be less sensitive for women with dense breasts [7]. The DENSE trial [8] has studied the value of MRI screening on women in the highest density category (ACR category 4) and has shown its effect on decreasing interval cancer rate.

However, due to the high dimensionality of the breast DCE-MRI data, interpretation is time-consuming and complex. Therefore, computer-aided detection (CADe) systems have been developed to help and support radiologists to analyze breast MRI with the goal to reduce interpretation time and reduce oversight errors. CADe systems automatically mark the breast's most suspicious locations and assist the radiologist in detecting a lesion that might be overlooked or misinterpreted [9,10]. A previous study [11] on sixteen positive and prior-negative MRI pair exams has shown that more than half of the MRI detected cancers could be detected on prior MRI scans. Two other studies [9,12] have also shown that the CADe system can detect several overlooked or misinterpreted cancers and, therefore, can reduce oversight errors. Furthermore, fully automatic CADe systems have the potential to decrease intra and inter-observer variability [13].

Traditionally, most current breast MRI CADe systems detect lesions based on morphological characteristics and dynamic information obtained from the late-phase scans, such as the wash-out of contrast agents, which requires the acquisition of several volumes up to about 7 minutes after contrast administration [14]. A study has shown that kinetic features extracted from the time-signal intensity curve in a full dynamic MRI protocol are suitable for lesion detection. The 3D morphological features are effectively helpful in reducing false positives [15]. In this sense, some CADe systems work in opposed direction as most breast radiologists who tend to assess lesions based on morphology first, using the kinetic curve to increase the specificity.

The TWIST sequence (time-resolved angiography with interleaved stochastic trajectories) [16], used in this study, is an ultrafast view-sharing MRI technique providing a high temporal resolution while retaining a high spatial resolution. This combination makes it possible to evaluate both the morphological and dynamic information [17,18]. In the clinical MRI protocol, 20 TWIST acquisitions are acquired during the inflow of contrast, thus providing early-phase dynamic contrast uptake information; hence, it considers the contrast wash-in rather than contrast wash-out.

In contrast to conventional computer vision methods that use handcrafted features to detect lesions in breast DCE-MRI [9,15], deep-learning techniques learn these features directly from the data [19]. Utilizing deep-learning-based approaches to analyze and evaluate breast DCE-MRI has shown promising results [12,14,20]. It is to be expected that deep-learning-based algorithms will play an important role in facilitating the widespread use of MRI as an economical and suitable screening method that can support the radiologists' decisions and bring their performances closer together [3].

This study proposes a CADe method based on a deep learning-based one-stage detector to detect breast lesions in ultrafast DCE-MRI by exploiting both spatial and temporal information obtained from the early phase of contrast enhancement. In our method, MRI images of the breast are used as input for a 3D RetinaNet model after motion compensation and temporal alignment using registration to localize the lesion bounding boxes. Using bounding boxes rather than segmentation of the lesions allows for faster annotation of the dataset. Besides this, segmenting lesions typically has to be done on the subtraction volumes, enabling bias both by registration and contrast leakage in the normal surrounding parenchyma. The original RetinaNet model, designed for 2D images, is modified to detect 3D breast lesions, which typically only occupy a small region in the whole breast. Using focal loss, the model focuses on detecting less visible lesions, which leads to an improvement in detecting cancerous lesions. To evaluate the model, we employed 10-fold cross-testing at the patient level to avoid bias in our assessment.

The main contributions of this paper are described as follows. The proposed method utilizes TWIST sequences as an ultrafast DCE-MRI protocol previously used in computer-aided diagnosis (CADx) methods [21]; however, to our knowledge, it is the first time this has been utilized in a deep learning-based CADe method for breast lesions. The RetinaNet model is modified to adapt to 4D TWIST data and localize lesions that mostly occupy a small area inside the breast. Both 3D spatial information and temporal information are available in the proposed method for lesion detection. Benign lesions that are less visible, proven by biopsy or clinical follow-up, are added to the method as hard examples to improve the CADe performance in localizing lesions.

This paper is organized as follows. In Section 2, the proposed CADe method is introduced, and also the breast MRI data acquisition is described in detail. Experimental results are presented and discussed in Section 3 and 4, respectively, and Section 5 concludes the paper.

## 2 Materials and Methods

### 2.1 Breast MRI Data Acquisition

The dataset used in this study is similar to one utilized in our previous work for breast lesion classification [14]. The dataset was composed of breast MRI scans collected at the Radboud university medical center radiology department, Nijmegen, the Netherlands, between October 2011 and December 2016, by retrospective evaluation of all consecutive cases with a clinical indication for breast MRI. The institutional review board waived the need for informed consent. Breast MRI was performed at our institute for the screening of women with an intermediate or high risk of developing breast cancer and preoperative staging in women with invasive carcinoma, including women with invasive lobular carcinoma (ILC) , those under 50 years old, or those with uncertain tumor size as well as the assessment of women selected for neoadjuvant chemotherapy, with lymph node metastasis from an unknown source, and with unsolvable findings from other imaging modalities.

All DCE-MRIs were acquired in axial orientation on a 3T MR scanner (Siemens Magnetom Trio/Skyra; Erlangen, Germany) with a dedicated 16-channel bilateral breast coil (Siemens) using a bi-temporal protocol. In the T1-weighted (T1w) DCE-MRI acquisitions process, a series of twenty ultrafast TWIST acquisitions with 4.3 s intervals were acquired during the inflow of contrast agent, and conventional DCE-MRI series were also obtained during the late phase of contrast enhancement. The contrast agent was injected at a dose of 0.1 mmol/kg (Dotarem, Guerbet, France) using a power injector (Medrad, Warrendale, PA, USA) at a flow rate of 2.5 mL/s, followed by a saline flush. Only ultrafast TWIST series without fat suppression were investigated in this study. The Matrix size of TWIST volumes was 384×384×60 resulting in a spatial resolution of 1.0×0.9×2.5 mm$^3$. Other clinical imaging parameters are as follows: field of view (FOV) = 360 mm, flip angle (FA) = 20°, repetition time (TR) = 3.96 ms and echo time (TE) = 2.2 ms.

The dataset contained 572 lesions from 489 breast DCE-MRI studies on 462 individual patients between 21 and 89 years old (mean 50 years old). There were 365 malignant lesions proven by histopathology with an average effective radius of 13.6 mm and a standard deviation of 7.5 mm. In total, there were 207 benign lesions included 148 lesions proven by histopathology with an average effective radius of 12.0 mm and a standard deviation of 7.5 mm and 59 non-biopsied lesions reported by the radiologist and supported by at least two-years clinical follow-up with an average effective radius of 8.3 mm and a standard deviation of 3.5 mm. Based on the pathological information available for most of the lesions, there were 215 invasive ductal carcinomas, 78 invasive lobular carcinomas, 45 pure in situ ductal carcinomas amongst the malignant lesions, and 39 fibroadenomas amongst biopsied benign lesions, and the rest considered various pathological diagnoses. Also, due to the radiological report, the dataset contained both mass and non-mass lesions.

All lesions were manually annotated on TWIST subtraction volumes by indicating two points describing a 3D circumscribed bounding box of the lesion under an expert breast radiologist's supervision. By assessing the dimensions of the bounding boxes, half of the benign lesions and half of the malignant lesions had the same range of volume between 3 cm$^3$ and 16 cm$^3$, indicating that overlap between the size of the benign and malignant lesions is relatively considerable.

## 2.2 Preprocessing

An overview of the proposed CADe method included preprocessing techniques, and a modified 3D RetinaNet model used for automatic lesion detection is shown in Fig. 1.

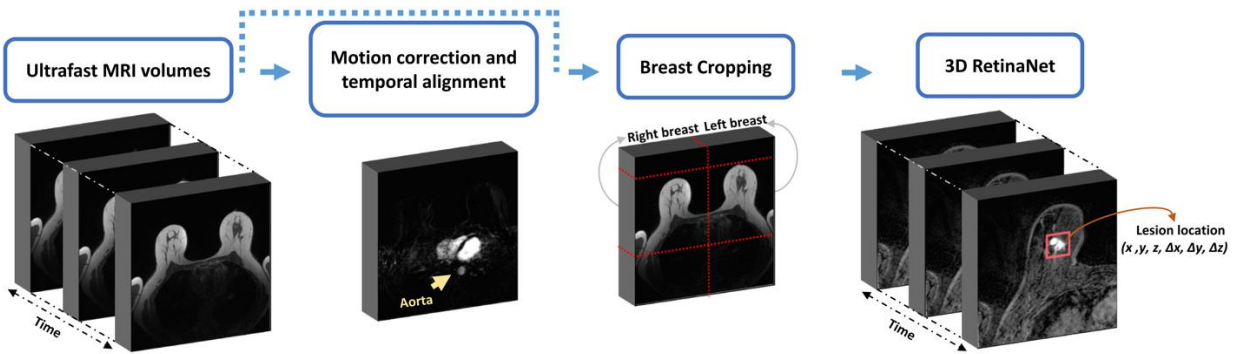

**Fig. 1** Overview of the proposed CADe method

Even though the TWIST series are approximately robust against motion since it is a view-sharing algorithm, motion compensation is still useful for lesion detection. Motion artifacts are unavoidable due to respiration, cardiac motion, pectoral muscle relaxation, and other motion sources during the image acquisition process. They can influence lesion's kinetic and morphological characteristics and increase the false positives in breast DCE-MRI [22].

Motion compensation is done by combining rigid and non-rigid transformations in a three-resolution scheme to register all TWIST volumes to the first volume acquired before contrast agent administration. A rigid transformation containing translation and rotation is applied for global alignment in 250 iterations, followed by a B-spline non-rigid transformation in 500 iterations to minimize the local variations between two volumes. Adaptive stochastic gradient descent optimizer and the mutual information similarity measure are used in the registration process. The algorithm has been implemented using Elastix version 4.8 [23].

The subtracted volumes are then obtained by subtracting the pre-contrast volume from all registered volumes. Due to the variations in injection time or blood circulation for each patient, the arrival of contrast within the breast is variable between patients. Therefore, the time-point at which the descending aorta is visible for the first time due to the contrast agent administration is considered the reference time-point [18,21], and it is assumed that earlier time-points do not add to the lesion detection capabilities of the algorithm. Thirteen post-contrast volumes, starting from the reference time-point, are utilized for lesion detection because this is the maximum available number of subtracted volumes for all patients. In other words, all the patients' images are normalized based on the time when the contrast agent reached the descending aorta, which is estimated to be very close to the start of enhancement within the breast [14].

Breasts are cropped to exclude the air around and the area behind the chest to decrease false-positive findings, focus on the breast tissue, reduce the model's input size, and reduce computation costs. As breasts should be centered during acquisition, left and right breasts are divided by halving the images equally. The ultrafast TWIST series are acquired without fat suppression; therefore, a simple Otsu thresholding method on the pre-contrast volume is used to segment the breast and indicate its top-point. The breast is cropped from five pixels above the top-point by 192 pixels. Consequently, a breast image with a size of 13×192×192×60 is provided for the model's input.

*2.3 Automated Breast Lesion Detection*

The RetinaNet architecture [24] is a deep learning-based one-stage detector employed in the proposed CADe to localize breast lesions. RetinaNet was developed for object detection on 2D natural images to improve accuracy while maintaining the speed and simplicity in one-stage detectors. In the proposed CADe, a modified 3D RetinaNet model is exploited to detect breast lesions in ultrafast DCE-MRI. The model structure can be seen in Fig. 2. 4D cropped breast images are fed into the model in which the temporal information is the channel dimension. Therefore, spatial information, such as 3D morphological features, and temporal information, such as kinetic features, are accessible for lesion detection. The lesion can be localized by learning the ground-truth bounding boxes by indicating six points in a three-dimensional space.

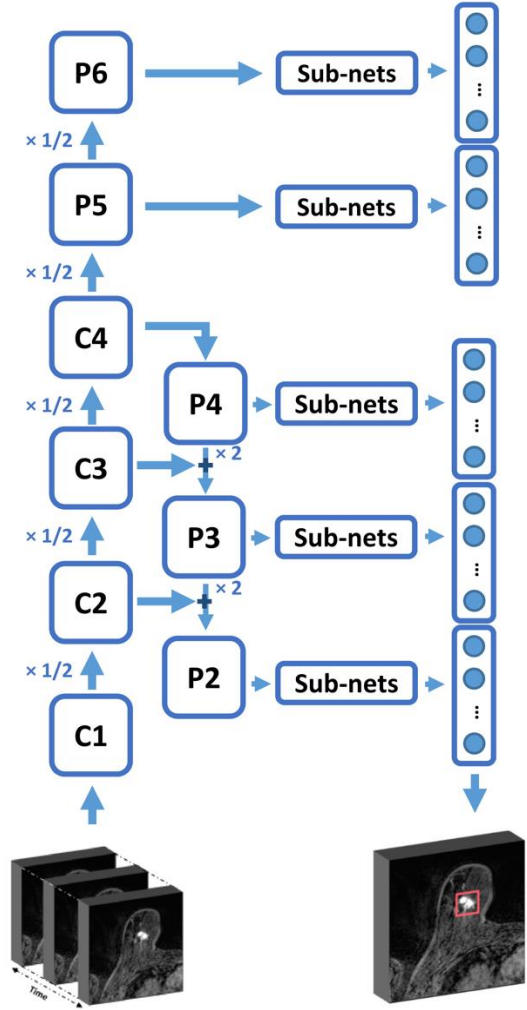

**Fig. 2** Modified 3D RetinaNet structure

RetinaNet's backbone network is a combination of a feature pyramid network (FPN) [25], which is a fully convolutional network (FCN) [26], and ResNet [27]. The FPN consists of vertical and lateral pathways converting the input image to a feature pyramid with several resolutions; hence breast lesions with various sizes could be detected by different pyramid levels. FPN is built on top of a 3D ResNet architecture. According to Fig. 2, {C1, C2, C3, C4} are outputs of the ResNet's building blocks in which the spatial resolution is decreased by a factor of 2 compared to the original image. These outputs generate a final feature pyramid with levels of P2-P6. The details of the pyramid levels and building blocks are available in [25,27].

The used structure has several differences compared to the original one, which is explained below. First, a 3D structure is used instead of the 2D structure, and the model details and layers are modified to adapt to the higher dimensional inputs. Second, the P1 pyramid level is still not used due to its high memory footprint; however, as the breast lesions are usually small compared to the whole breast volume, the P2 pyramid level with a high resolution is used for detecting small lesions. Third, the last building block of the ResNet (C5) is not used in this work, and P5 is alternatively computed by applying 3×3×3 stride-2 convolution on C4. ReLU is then computed from P5 output, and P6 is obtained by applying a 3×3×3 stride-2 convolution to that. P5 and P6 are mostly useful for detecting bulky lesions that rarely occur in screening; therefore, discarding C5 and computing P5 and P6 by a single convolutional layer still allows the detection of bulky lesions, while reducing computational costs, and increasing the speed. Moreover, the trilinear method is used for upsampling the feature maps by a factor of 2 through top-down pathways, and the final feature pyramid is attained in 256 channels for each layer.

Nine 3D anchor boxes are generated from every voxel of feature pyramid levels, which are empirically set based on the area covered by ground-truth bounding boxes in three aspect ratios. Anchor boxes of every pyramid level are fed into the two FCN subnets. The classification subnet predicts the probability of the lesion presence in each anchor box, and the regression subnet regresses the offset from the anchor box and ground-truth. These two separated subnets consist of four layers of 3×3×3 convolution, and the final outputs are reduced to 64 feature maps to decrease GPU memory consumption. Other details are as [24].

Focal loss [24] is computed on the classification subnet's output and incorporates the modulating factor and balancing weight factor to reshape the standard cross-entropy loss. The modulating factor is a dynamic weight factor that automatically down-weights the loss assigned to easy examples, which are well-classified and alternatively focusses on hard examples. Therefore, easy examples such as the vast majority of background examples or visible lesions which can be easily classified (e.g., most of BI-RADS 5 malignant lesions) could not significantly influence the loss, and the model focusses on learning hard examples such as differentiating less visible lesions from confounding structures inside the breast like vessels and fibroglandular tissue. For instance, detecting BI-RADS 2 benign lesions, which were proven without performing a biopsy and are not merely distinguishable from background, can be considered as hard examples for the model.

A balancing weight factor is added to reduce further the impact of background examples. As most breast MRIs are normal or only contain one lesion, a 3D breast image mostly includes background and non-lesion examples. Therefore, even if a small value of the loss is assigned to these examples, the sum of the losses will increase and affect the learning process. Consequently, this factor balances the importance of lesion and non-lesion examples.

The standard smooth L1 loss computed in the regression subnet is added to the focal loss to form the final loss. The anchors, with their intersection over union (IoU) with the lesion bounding boxes above the threshold of 0.2, are considered in loss computation. This smaller threshold is selected due to the less overlap between 3D bounding boxes and also imperfect ground-truth bounding boxes drawn for some examples. The model is trained with Adam optimizer with an initial learning rate of $10^{-4}$, which is decreased by 10 if the loss does not change for three consecutive epochs. Moreover, each batch contains eight breasts, and the model is not pre-trained. Other parameters are as in the original RetinaNet paper [24].

*2.4 Experiments and Evaluation Methods*

10-fold cross-testing is performed at the patient level to test the proposed CADe method. The dataset contains 59 benign lesions proven by clinical follow-up rather than biopsy. These are considered hard examples for the model because it is hard to distinguish them from the background. Approximately the same number of malignant, benign, and non-biopsied benign lesions are randomly assigned to each fold. Each patient's left and right breast are considered different data points but fed into the model in a single batch. Also, to assess the proposed method by a relatively independent test process which is more similar to a real situation, train and test sets are selected based on acquisition date. Therefore, the scans acquired between 2011 and 2014 are used as train and validation sets and more recent data are utilized for test set.

To evaluate the proposed method's performance, we compute the detection rate, sensitivity, detection rate of benign lesions, and computation performance metric (CPM). The detection rate is defined as the number of lesions detected by the proposed CADe (both benign and malignant) to all annotated lesions. We calculate the sensitivity as the fraction of malignant lesions detected divided by the total number of proven malignant lesions in the dataset. Since the proposed CADe method is designed to localize both benign and malignant lesions, we report a detection rate for benign lesions, as well. As some of the benign lesions are harder to be distinguished from the background in comparison to the malignant lesions, it is expected that this detection rate is lower. Furthermore, CPM is defined as the average of each evaluation metric at {1/8, 1/4, 1/2, 1, 2, 4, 8} false positives per normal breast [12]. Free-response operating characteristic (FROC) analysis is also used to assess the results. A lesion is considered as a detected lesion if intersection between the detected bounding box which is the output of the model and the ground-truth bounding box which is specified by the radiologist is above a certain threshold. False-positive findings are assumed the regions that do not contain annotated lesions. Due to the lack of normal cases in this study, false-positive findings are computed on breasts with no lesions because the breast, which contains the lesion, could include other suspicious regions that are related to the lesion but not annotated. Consequently, performance metrics are calculated at false-positive findings per normal breast. It should be realized that also breasts that we considered as normal may contain unreported lesions because not all lesions are in practice called by radiologists, which may have led to a higher number of false-positive findings in our results.

To perform statistical comparison between two FROC curves, p value is calculated as explained in [12]. The cases are sampled 1000 times, and FROC curves are plotted for Two comparative situations based on these samples. The p value is defined as the ratio of the number of the negative or zero-valued ΔCPM's (difference between two CPM values) to the number of samples. There is a significant difference between the two CPM values when the p value is ≤0.05.

## 3 Results

In this section, the proposed CADe method for detecting lesions in ultrafast DCE-MRI is reported and compared to the most recent methods.

### 3.1 Detection Results

The detection results for correctly detected, missed lesions, and false-positive findings in a single slice of a breast at a specified time-point are visually provided in Fig 3. The results have been achieved at 4 false positives per normal breast. As can be seen, the missed lesion is a non-biopsied one diagnosed by clinical follow-up and considered as a hard example that is difficult to differentiate from the background. False-positive findings are may also be caused by noise, motion artifacts, confounding structures such as fibroglandular tissue, skin, and vessels.

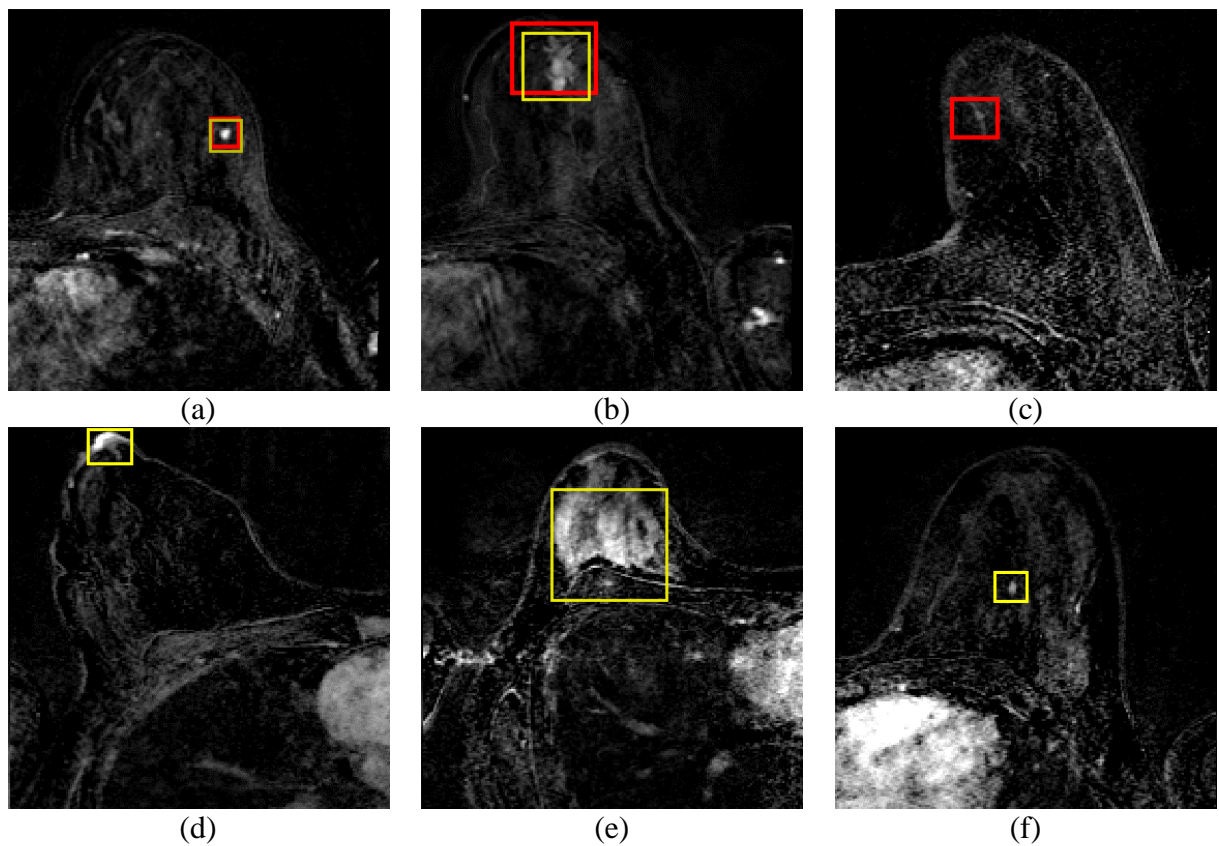

**Fig. 3** Detection results for a detected benign lesion (a), a detected malignant lesion (b), a missed benign lesion proven by the clinical follow-up (c), and false-positive examples: nipple (d), background enhancement of the normal fibroglandular tissue (e), an unreported lesion which detected by the model but considered as false-positive by mistake (f) shown in a 2D slice of the breast. red boxes sketch ground-truth bounding boxes drawn by the radiologist and yellow boxes sketch CADe outputs such as detected suspicious region or false-positive findings

FROC analysis for all lesions, malignant lesions, and benign ones is separately shown in Fig. 4. The results have been achieved by a 14-layers ResNet backbone network (ResNet-18 without last building block). An overall detection rate of 0.87 (0.840-0.918) and 0.90 (0.876-0.934), and a sensitivity for malignant lesions of 0.94 (0.903-0.973) and 0.95 (0.934-0.980) were obtained at

2 and 4 false positives per normal breast by using 10-fold cross-testing, respectively. The CPM for detection rate, sensitivity, and detection rate of benign lesions was 0.78, 0.86, and 0.64, respectively. Furthermore, by using the test process based on the acquisition date, an overall detection rate and sensitivity of 0.86 (0.823-0.907) and 0.93 (0.907-0.972) were obtained at 2 false positives per normal breast, respectively.

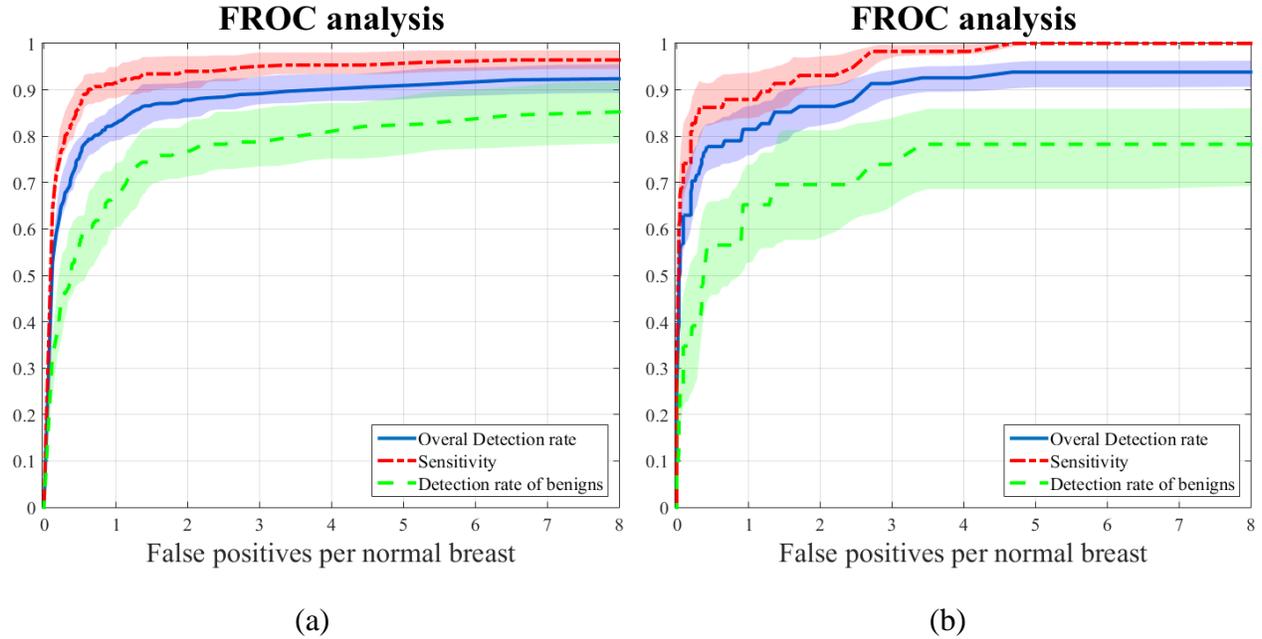

**Fig. 4** FROC curves of the proposed CADe method for all lesions (detection rate), malignant lesions (sensitivity), and benign lesions (detection rate of benign lesions) by using 10-fold cross-testing (a) and testing based on the acquisition date (b). The upper and lower boundaries of 95% confidence intervals are also plotted.

To analyze the influence of age on the performance of the proposed CADe method, the age of the cases with benign or malignant lesions and the cases with only malignant lesions that the proposed CADe detects, were investigated at two different false positive rates. The average age of the first group at 2 and 4 false positives per normal breast was 50.7 and 50.6 years old respectively, and for the second group was 53.0 for both rates. These are approximately equal to the average age of the whole cases (50 years old) and the average age of the cases with only malignant lesions (53 years old) and show the age does not influence the performance of the CADe method.

*3.2 Implementation Parameters*

One of the issues that should be investigated for implementing the proposed method is the depth of the backbone network. For this purpose, Fig 5(a) draws the FROC curve of the FPN-ResNet-18 and FPN-ResNet-50 backbone network on one of the most challenging folds of the dataset. It should be noticed that the ResNet in these backbone networks have 14 and 41 layers, respectively, due to the discard of the final building block. FROC curves plotted in Fig. 5(b) compare the impact of using different epochs in the training stage to achieve optimum performance empirically.

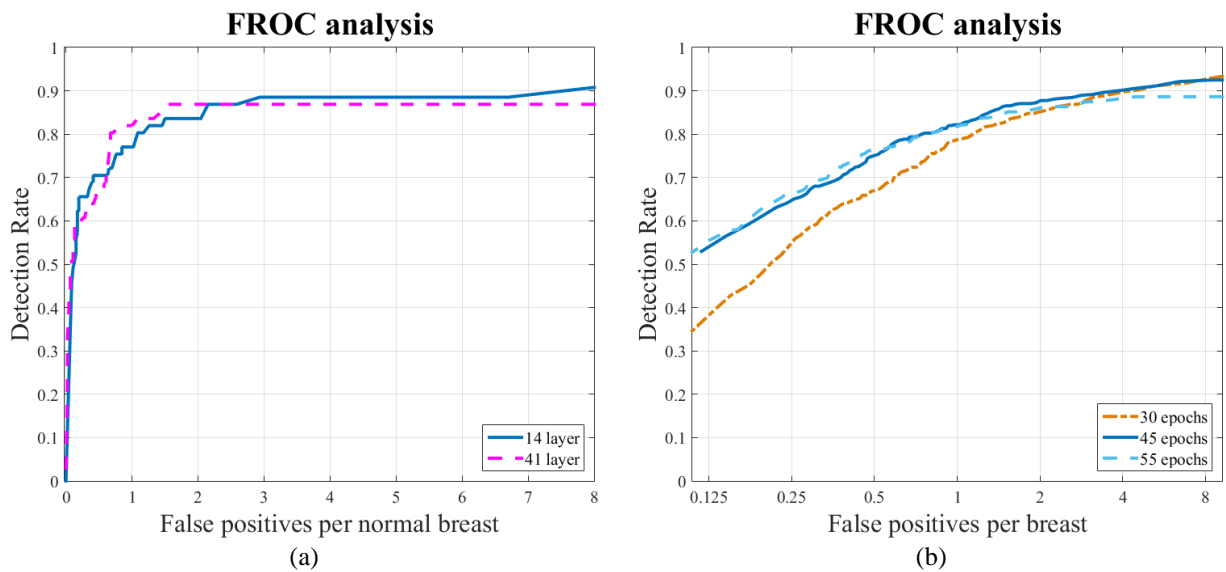

**Fig. 5** Comparison of performance of the proposed CADe method for different depths in the backbone network tested on a challenging fold of the dataset (a), different epochs in the training stage (b)

*3.3 Hard Example Effect*

To investigate the influence of using benign lesions (including the examples that are harder to be detected) on detecting cancerous tumors, we compared the performance of the proposed CADe with and without benign lesions (see Fig. 6). False-positive findings are still the regions that do not contain annotated lesions. The number of epochs has been set to achieve the best results for the without-benign scheme (i.e., the sensitivity for malignant lesions).

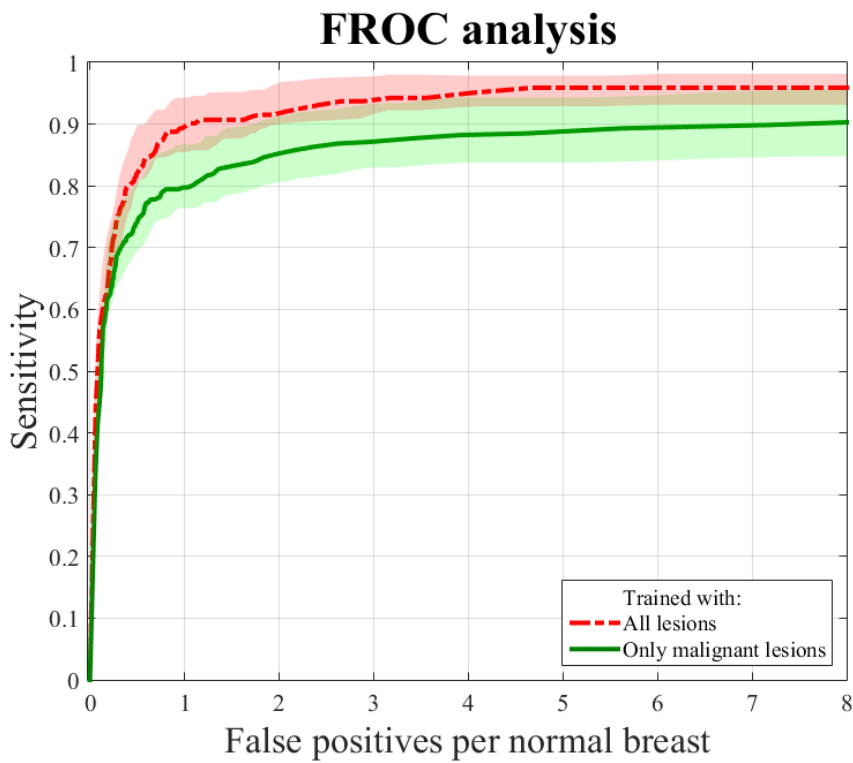

**Fig. 6** Comparison of FROC curves of the proposed CADe method for detecting cancerous tumors by using all lesions, including benign and malignant lesions and using only malignant lesions. The upper and lower boundaries of 95% confidence intervals are also plotted.

*3.4 Conventional DCE-MRI*

The proposed deep learning-based model has been tuned for ultrafast DCE-MRI. Fig.7 compares the performance of the proposed CADe system on Conventional and ultrafast DCE-MRI. The FROC curves for the sensitivity measure have been achieved by using subtracted volume and the same model architecture, parameters and preprocessing steps designed for ultrafast sequences. The calculated p value for the statistical comparison between the FROC curves was 0.032 which means that the differences between CPM values by using ultrafast and conventional DCE-MRI is statistically significant.

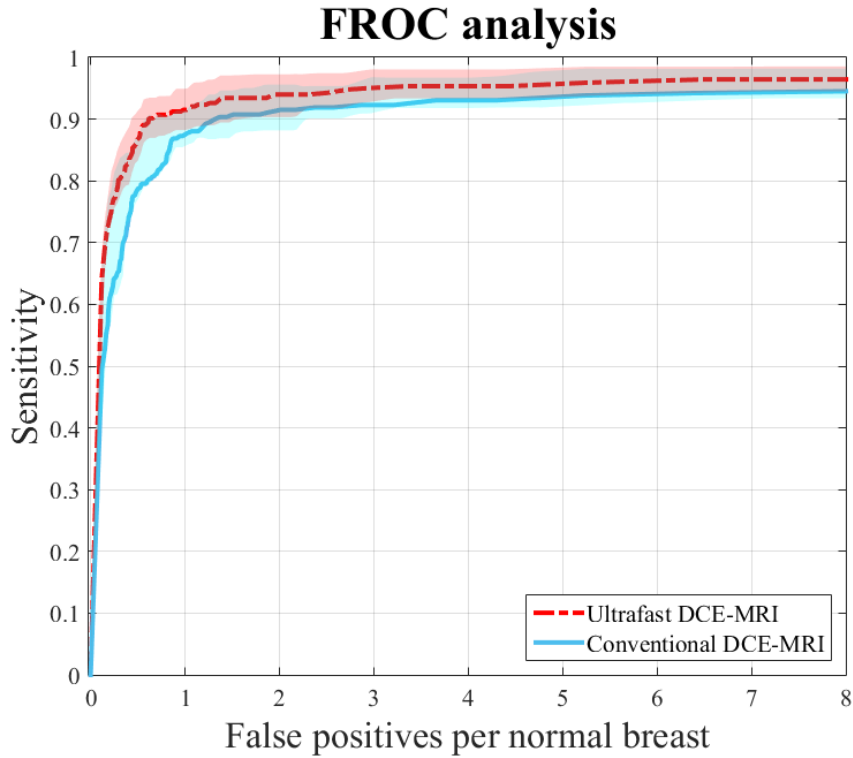

**Fig. 7** Comparison of FROC curves for the proposed CADe method on ultrafast and conventional DCE-MRI. The upper and lower boundaries of 95% confidence intervals are also plotted.

### 3.5 Comparison with State-of-the-Art Methods

Due to the lack of standard publically available datasets for breast DCE-MRI and different evaluation criteria utilized by studies, comparing the CADe methods is complicated and inherently incomplete. However, Table 1 compares the performance of the proposed method and the state-of-the-art methods by various metrics and mentions the employed datasets' size.

Table 1 Performance comparison between the other methods and the proposed method

| Methods | Evaluation Criteria | | | False Positives | Number of lesions | | |
|---|---|---|---|---|---|---|---|
| | Det. | Sens. | CPM | | Total | Benign | Malignant |
| Proposed method | 0.87 | 0.94 | 0.86 | 2/ normal breast | 572 | 207 | 365 |
| Gubern-Mérida et al. [9] | - | 0.89 | - | 4/ normal case | 105 | - | 105 |
| Gubern-Mérida et al. [10] | - | 0.82 | 0.53 | 4/ normal case | 42 | - | 42 |
| Dalmiş et al. [12] | - | - | 0.64 | - | 42 | - | 42 |
| Vignati et al. [13] | 0.89 | - | - | 4/ breast | 65 | 12 | 53 |
| Chang et al. [15] | 0.93 | - | - | 6.15/ case | 95 | 28 | 67 |
| Maicas et al. [20] | 0.8 | - | - | 3.2/ case | 69 | 23 | 46 |
| Shokouhi et al. [22] | 0.94 | - | - | 5.29/case | 170 | 63 | 107 |
| McClymont et al. [28] | 1.0 | - | - | 4.5/ case | 93 | 35 | 58 |
| Renz et al. [29] | - | 0.96 | - | 0.16/case | 141 | 53 | 88 |
| Maicas et al. [30] | 0.8 | - | - | 2.8/ case | 69 | 23 | 46 |
| Amit et al. [31] | - | 0.98 | - | 7/ case | - | - | - |

Det: Detection Rate, Sens: Sensitivity

## 4 Discussion

We built a CADe system using a relatively large dataset of breast DCE-MRI scans containing a realistic mixture of malignant and benign lesions. In this dataset, the CADe system achieved a relatively high sensitivity of 0.94 (0.903-0.973) at 2 false positives per normal breast. Two different test processes were applied and achieved almost the same performance that provided an insight into the generalizability of the proposed CADe method. As can be seen from the visible results provided in Fig. 3, the proposed CADe method detects the lesions by localizing a bounding box instead of the segmentation. Therefore, the annotation of the dataset becomes simpler in comparison with defining the lesion borders. Moreover, accurate segmentation can be quickly done by a region-growing algorithm within the detected bounding boxes. For radiologists and several computer-aided diagnosis (CADx) methods that differentiate benign and malignant lesions [14, 32], a detected bounding box is sufficient.

The proposed CADe method localizes both benign and malignant lesions. It is a complement and prerequisite for CADx systems such as the one introduced in [14] applied on the same dataset. Therefore, the first step is to localize the lesions regardless of their type, then the lesions can be classified to benign and malignant by the CADx system. As a result, we can have a fully automated system by combining the proposed CADe method and the existing CADx for breast cancer diagnosis on high-dimensional MRI data. Also, the normal breast can help the model to learned the general shape of the breast and structures inside it to distinguish them from the lesions.

In addition to the importance of detecting malignant lesions, detecting the annotated benign lesions and distinguishing them from the background and confounding structures inside the breast like vessels, skin, and fibroglandular tissue proved to be beneficial for the system's performance (Fig. 6). While the overall detection rate of benign lesions remained lower, this is not considered a problem in the clinic, as benign lesions would be disregarded when they are not suspicious and can be distinguished by a CADx method later.

Fig. 5(a) shows that increasing the backbone network's depth does not improve the performance and reduces the speed. Deeper networks with more complex layers are more suitable for object detection in natural images due to their complex hierarchical relations, but this is not the case for breast DCE-MRI.

According to Fig. 5(b), using 45 epochs achieves the best performance, while a higher or lower number of epochs cause overfitting and underfitting, respectively. RetinaNet, as a one-stage detector, provides reasonable accuracy and speed simultaneously and uses less GPU memory than deep learning-based segmentation methods such as U-Net [33].

Ultrafast DCE-MRI protocol provides early-phase information compared to full dynamic MRI protocol as well as temporal information. As shown in Fig.7, the performance of the proposed CADe method by using ultrafast DCE-MRI is higher than its performance by using conventional DCE-MRI, which is also proved by statistical comparison. Consequently, using a RetinaNet architecture-based CADe system could be helpful for breast lesion detection in ultrafast DCE-MRI. It should be noticed that to improve the model performance on conventional DCE-MRI, the preprocessing steps, and model parameters and input could be modified based on this data. Besides the other common MRI sequences, such as DWI, T2w, and T1w, the ultrafast sequences could be acquired during the T1w DCE-MRI acquisitions process. Therefore, by using the combination of the proposed method and methods designed on the other available MRI sequences, a performance improvement can be expected.

Table 1 compares the proposed CADe and current state-of-the-art methods. However, it is not a thorough comparison because the different datasets with a different distribution of types and number of lesions and different MRI acquisition parameters were used in the studies. Evaluation criteria were also different, some methods detected both benign and malignant lesions and reported detection rate, but some others investigated cancerous lesion detection alone, reporting sensitivity. These criteria were obtained at various false positives rates, which make them harder to be compared. Furthermore, methods such as those presented in Renz et al. [29] and Chang et al. [15] did not present FROC analysis. It should be noticed that the proposed method is the only one investigating ultrafast DCE-MRI and has the most extensive dataset. Although, the dataset, like many other methods, does not only include the screening data, the model has been designed to emphasize the detection of small lesions that are more common in the screening setting.

CADe methods can be divided into three categories; First, image processing and clustering-based methods. McClymont et al. [28] achieved the high result by combining mean-shift clustering and graph-cuts-based method, but they used a medium-size multi-modal dataset. Renz et al. [29] exploited adaptive thresholding to obtain high sensitivity; however, ground-truth annotations were not provided in their method, and the evaluation was performed by visual inspection. Shokouhi et al. [22] applied fuzzy c-means clustering, and their system performance was reported at a high false positive rate. Vignati et al. [13] utilized a normalization-based method and introduced one of the primary CADe methods.

Second, learning-based methods which detect the lesions in a supervised classification scheme. Gubern-Mérida et al. [9] and Gubern-Mérida et al. [10] used various kinetic and morphological features and reported their results at different false-positive measure. Also, Chang et al. [15] reported their results at a high false positive rate.

Third, deep learning-based methods. Maicas et al. [20] and Maicas et al. [30] used deep learning to speed up lesion detection, and Dalmiş et al. [12] detected the lesions in an abbreviated MRI protocol and used CPM for reporting the performance. Amit et al. [31] employed deep learning, but the lesions were detected on selected 2D breast slices for each case.

A limitation of the study was that we only used cases with lesions, and no normal cases were involved in our study, whereas adding the normal cases to the dataset can make it more thorough.

## 5   Conclusion

This paper proposed a new CADe method based on a 3D RetinaNet model, a deep learning-based one-stage detector. The proposed method was applied to breast ultrafast DCE-MRI sequences to localize the lesions by considering both 3D morphological and dynamic information obtained from early-phase scans. Our method achieved promising results and could support radiologists in analyzing ultrafast DCE-MRI data by marking suspicious regions.

*Ethical approval and Informed consent*

Ethics approval is obtained from the institutional review board. The institutional review board waived the need for informed consent.

*Conflict of Interest*

The authors have no relevant conflicts of interest to disclose.

*Data Availability Statement*

Research data are not shared.

*References*

1. Fusco R, Sansone M, Filice S, et al. Pattern Recognition Approaches for Breast Cancer DCE-MRI Classification: A Systematic Review. *Journal of Medical and Biological Engineering*. 2016;36(4):449-459.
2. Siegel RL, Miller KD, Jemal A. Cancer statistics, 2016. *CA: A Cancer Journal for Clinicians*. 2016;66(1):7-30.
3. Teuwen J, Mann R, Moriakov N. AI applications in breast imaging. *Artificial Hype*. 2019;19(2)
4. Saslow D, Boetes C, Burke W, et al. American Cancer Society guidelines for breast screening with MRI as an adjunct to mammography. *CA Cancer J Clin*. 2007;57(2):75-89.
5. Vreemann S, Gubern-Mérida A, Schlooz-Vries MS, et al. Influence of Risk Category and Screening Round on the Performance of an MR Imaging and Mammography Screening Program in Carriers of the BRCA Mutation and Other Women at Increased Risk. *Radiology*. 2018;286(2):443-451.

6. Vreemann S, van Zelst JCM, Schlooz-Vries M, et al. The added value of mammography in different age-groups of women with and without BRCA mutation screened with breast MRI. *Breast Cancer Res*. 2018;20(1):84.

7. Wanders JO, Holland K, Veldhuis WB, et al. Volumetric breast density affects performance of digital screening mammography. *Breast Cancer Research and Treatment*. 2017;162(1):95-103.

8. Emaus MJ, Bakker MF, Peeters PH, et al. MR Imaging as an Additional Screening Modality for the Detection of Breast Cancer in Women Aged 50-75 Years with Extremely Dense Breasts: The DENSE Trial Study Design. *Radiology*. 2015;277(2):527-537.

9. Gubern-Mérida A, Martí R, Melendez J, et al. Automated localization of breast cancer in DCE-MRI. *Medical Image Analysis*. 2015;20(1):265-274.

10. Gubern-Mérida A, Vreemann S, Martí R, et al. Automated detection of breast cancer in false-negative screening MRI studies from women at increased risk. *European Journal of Radiology*. 2016;85(2):472-479.

11. Yamaguchi K, Schacht D, Newstead GM, et al. Breast cancer detected on an incident (second or subsequent) round of screening MRI: MRI features of false-negative cases. *AJR. American Journal of Roentgenology*. 2013;201(5):1155-1163.

12. Dalmış MU, Vreemann S, Kooi T, Mann RM, Karssemeijer N, Gubern-Mérida A. Fully automated detection of breast cancer in screening MRI using convolutional neural networks. *Journal of Medical Imaging (Bellingham)*. 2018;5(1):014502.

13. Vignati A, Giannini V, De Luca M, et al. Performance of a fully automatic lesion detection system for breast DCE-MRI. *Journal of Magnetic Resonance Imaging*. 2011;34(6):1341-1351.

14. Dalmiş MU, Gubern-Mérida A, Vreemann S, et al. Artificial Intelligence-Based Classification of Breast Lesions Imaged With a Multiparametric Breast MRI Protocol With Ultrafast DCE-MRI, T2, and DWI. *Investigative Radiology*. 2019;54(6):325-332.

15. Chang YC, Huang YH, Huang CS, Chen JH, Chang RF. Computerized breast lesions detection using kinetic and morphologic analysis for dynamic contrast-enhanced MRI. *Magnetic Resonance Imaging*. 2014;32(5):514-522.

16. Laub G, Kroeker R. Syngo TWIST for dynamic time-resolved MR angiography. *Magnetom Flash*. 2006;34:92–95.

17. Milenković J, Dalmış MU, Žgajnar J, Platel B. Textural analysis of early-phase spatiotemporal changes in contrast enhancement of breast lesions imaged with an ultrafast DCE-MRI protocol. *Medical Physics*. 2017;44(9):4652-4664.

18. Platel B, Mus R, Welte T, Karssemeijer N, Mann R. Automated characterization of breast lesions imaged with an ultrafast DCE-MR protocol. *IEEE Transactions on Medical Imaging*. 2014;33(2):225-232.

19. Zhang J, Saha A, Zhu Z, Mazurowski MA. Hierarchical Convolutional Neural Networks for Segmentation of Breast Tumors in MRI With Application to Radiogenomics. *IEEE Transactions on Medical Imaging.* 2019;38(2):435-447.

20. Maicas G, Carneiro G, Bradley AP, Nascimento JC, Reid I. Deep Reinforcement Learning for Active Breast Lesion Detection from DCE-MRI. *International Conference on Medical Image Computing and Computer Assisted Intervention − MICCAI 2017*. 2017;665-673.

21. Dalmış MU, Gubern-Mérida A, Vreemann S, Karssemeijer N, Mann R, Platel B. A computer-aided diagnosis system for breast DCE-MRI at high spatiotemporal resolution. *Medical Physics*. 2016;43(1):84.

22. Shokouhi SB, Fooladivanda A, Ahmadinejad N. Computer-aided detection of breast lesions in DCE-MRI using region growing based on fuzzy C-means clustering and vesselness filter. *EURASIP Journal on Advances in Signal Processing*. 2017;39.

23. Klein S, Staring M, Murphy K, Viergever MA, Pluim JP. elastix: a toolbox for intensity-based medical image registration. *IEEE Transactions on Medical Imaging*. 2010;29(1):196-205.

24. Lin TY, Goyal P, Girshick R, He K, Dollar P. Focal Loss for Dense Object Detection. *IEEE Transactions on Pattern Analysis and Machine Intelligence*. 2020;42(2):318-327.

25. Lin TY, Dollar P, Girshick R, He K, Hariharan B, Belongie S. Feature Pyramid Networks for Object Detection. *The IEEE Conference on Computer Vision and Pattern Recognition (CVPR)*. 2017;2117-2125.

26. Long J, Shelhamer E, Darrell T. Fully convolutional networks for semantic segmentation. *IEEE Conference on Computer Vision and Pattern Recognition (CVPR)*. 2015;3431-3440.

27. He K, Zhang X, Ren S, Sun J. Deep Residual Learning for Image Recognition. *IEEE Conference on Computer Vision and Pattern Recognition (CVPR),* 2016;770-778.

28. McClymont D, Mehnert A, Trakic A, Kennedy D, Crozier S. Fully automatic lesion segmentation in breast MRI using mean-shift and graph-cuts on a region adjacency graph. *Journal of Magnetic Resonance Imaging*. 2014;39(4):795-804.

29. Renz DM, Böttcher J, Diekmann F, et al. Detection and classification of contrast-enhancing masses by a fully automatic computer-assisted diagnosis system for breast MRI. *Journal of Magnetic Resonance Imaging*. 2012;35(5):1077-1088..

30. Maicas G, Carneiro G, Bradley AP. Globally optimal breast mass segmentation from DCE-MRI using deep semantic segmentation as shape prior. *IEEE 14th International Symposium on Biomedical Imaging (ISBI 2017)*. 2017;305-309.

31. Amit G, Hadad O, Alpert S, Tlusty T, Gur Y, Ben-Ari R, Hashoul S. Hybrid Mass Detection in Breast MRI Combining Unsupervised Saliency Analysis and Deep Learning. *International Conference on Medical Image Computing and Computer Assisted Intervention − MICCAI 2017*. 2017; 594-602

32. Ayatollahi F, Shokouhi SB, Teuwen J. Differentiating benign and malignant mass and non-mass lesions in breast DCE-MRI using normalized frequency-based features. *International Journal of Computer Assisted Radiology and Surgery*. 2020;15(2):297-307.

33. Ronneberger O, Fischer P, Brox T. U-Net: Convolutional Networks for Biomedical Image Segmentation. *International Conference on Medical Image Computing and Computer-Assisted Intervention – MICCAI 2015*. 2015;234-241.